\documentstyle[12pt]{article}
\begin{document}

\begin{flushright}{December, 2004}
\end{flushright}
\vskip 0.5 truecm

\begin{center}
{\Large{\bf  Topological properties of Berry's phase}}
\end{center}
\vskip .5 truecm
\centerline{\bf Kazuo Fujikawa }
\vskip .4 truecm
\centerline {\it Institute of Quantum Science, College of 
Science and Technology}
\centerline {\it Nihon University, Chiyoda-ku, Tokyo 101-8308, 
Japan}
\vskip 0.5 truecm

\begin{abstract}
By using a second quantized formulation of level crossing, which
does not assume adiabatic approximation, a convenient formula for
 geometric terms including off-diagonal terms is derived.
The analysis of geometric phases is reduced to a simple 
diagonalization of the Hamiltonian in the present formulation. 
If one diagonalizes the geometric terms in the infinitesimal 
neighborhood of level crossing, the geometric phases become 
trivial for any finite time interval $T$. The topological 
interpretation of Berry's phase such as the topological proof
of phase-change rule thus fails in the practical 
Born-Oppenheimer approximation, where a large but finite ratio
of two time scales is involved.
\end{abstract}


The geometric phases are mostly defined in the framework of 
adiabatic approximation~\cite{berry}-\cite{kuratsuji}, though
a non-adiabatic treatment has been considered in, for example, 
\cite{aharonov} and the (non-adiabatic) correction to the 
geometric phases has been analyzed~\cite{berry2}. One may then 
wonder if some of the characteristic properties generally 
attributed to the geometric
phases are the artifacts of the approximation. We here show 
that the topological properties of the geometric phases 
associated with level crossing are the artifacts of the adiabatic 
approximation which assumes the infinite time interval
$T\rightarrow \infty$~\cite{simon}. To substantiate this statement, we start
with  the exact definition of geometric terms associated with
level crossing. 
The level crossing problem is neatly formulated by using the 
second quantization technique without assuming 
adiabatic approximation. The analysis of phase factors in this 
formulation is reduced to a diagonalization of the Hamiltonian,
and thus it is formulated both in the path integral and in the 
operator formulation. We thus derive a convenient formula for 
geometric terms~\cite{berry} and their off-diagonal counter
parts.  (As for off-diagonal geometric phases, see  
\cite{manini} where the off-diagonal geometric phases in the 
framework of an adiabatic picture in the 
first quantization have been proposed, and their properties have
 been analyzed in~\cite{mukunda, hasegawa1, hasegawa2}.) Our 
formula allows us to analyze the topological 
properties of the geometric terms precisely in the infinitesimal 
neighborhood of level crossing. At the level crossing point,
the conventional energy eigenvalues become degenerate but the 
degeneracy is lifted if one diagonalizes the geometric terms.
It is then shown that the geometric phases become trivial (and 
thus no monopole singularity) in the infinitesimal neighborhood 
of level crossing for any finite time interval $T$. This is 
proved independently of the adiabatic approximation.  
The topological interpretation~\cite{stone, berry} of geometric 
phases such as the topological proof of Longuet-Higgins' 
phase-change rule~\cite{higgins} thus fails for any 
finite $T$ such as in the practical Born-Oppenheimer approximation, where
a large but finite ratio of two time scales is involved. 
In practical physical applications of geometric phases, finite 
$T$ is always relevant and our analysis implies that the widely 
used terminology of ``topological phases'' for the geometric 
phases should be taken with great care.

We start with the generic (hermitian) Hamiltonian 
\begin{equation}
\hat{H}=\hat{H}(\hat{\vec{p}},\hat{\vec{x}},X(t))
\end{equation}
for a single particle theory in a slowly varying background 
variable $X(t)=(X_{1}(t),X_{2}(t),...)$.
The path integral for this theory for the time interval
$0\leq t\leq T$ in the second quantized 
formulation is given by 
\begin{eqnarray}
Z&=&\int{\cal D}\psi^{\star}{\cal D}\psi
\exp\{\frac{i}{\hbar}\int_{0}^{T}dtd^{3}x[
\psi^{\star}(t,\vec{x})i\hbar\frac{\partial}{\partial t}
\psi(t,\vec{x})\nonumber\\
&&-\psi^{\star}(t,\vec{x})
\hat{H}(\hat{\vec{p}},\hat{\vec{x}},X(t))\psi(t,\vec{x})] \}.
\end{eqnarray}
We then define a complete set of eigenfunctions
\begin{eqnarray}
&&\hat{H}(\hat{\vec{p}},\hat{\vec{x}},X(0))u_{n}(\vec{x},X(0))
=\lambda_{n}u_{n}(\vec{x},X(0)), \nonumber\\
&&\int d^{3}xu_{n}^{\star}(\vec{x},X(0))u_{m}(\vec{x},X(0))=
\delta_{nm},
\end{eqnarray}
and expand 
$\psi(t,\vec{x})=\sum_{n}a_{n}(t)u_{n}(\vec{x},X(0))$.
We then have  $
{\cal D}\psi^{\star}{\cal D}\psi=\prod_{n}{\cal D}a_{n}^{\star}
{\cal D}a_{n}$
and the path integral is written as 
\begin{eqnarray}
Z&=&\int \prod_{n}{\cal D}a_{n}^{\star}
{\cal D}a_{n}
\exp\{\frac{i}{\hbar}\int_{0}^{T}dt[
\sum_{n}a_{n}^{\star}(t)i\hbar\frac{\partial}{\partial t}
a_{n}(t)\nonumber\\
&&-\sum_{n,m}a_{n}^{\star}(t)E_{nm}(X(t))a_{m}(t)] \}
\end{eqnarray}
where 
\begin{eqnarray}
E_{nm}(X(t))=\int d^{3}x u_{n}^{\star}(\vec{x},X(0))
\hat{H}(\hat{\vec{p}},\hat{\vec{x}},X(t))u_{m}(\vec{x},X(0))
.
\end{eqnarray}

We next perform a unitary transformation
$a_{n}=U(X(t))_{nm}b_{m}$
where 
\begin{eqnarray}
U(X(t))_{nm}=\int d^{3}x u^{\star}_{n}(\vec{x},X(0))
v_{m}(\vec{x},X(t))
\end{eqnarray}
with the instantaneous eigenfunctions of the Hamiltonian
\begin{eqnarray}
&&\hat{H}(\hat{\vec{p}},\hat{\vec{x}},X(t))v_{n}(\vec{x},X(t))
={\cal E}_{n}(X(t))v_{n}(\vec{x},X(t)), \nonumber\\
&&\int d^{3}x v^{\star}_{n}(\vec{x},X(t))v_{m}(\vec{x},X(t))
=\delta_{n,m}.
\end{eqnarray}
We emphasize that $U(X(t))$ is a unit matrix both at $t=0$ and 
$t=T$ if $X(T)=X(0)$, and thus $\{a_{n}\}=\{b_{n}\}$ both at 
$t=0$ and $t=T$. We take the time $T$ as the period of the 
slow variable $X(t)$.
We can thus re-write the path integral as 
\begin{eqnarray}
&&Z=\int \prod_{n}{\cal D}b_{n}^{\star}{\cal D}b_{n}
\exp\{\frac{i}{\hbar}\int_{0}^{T}dt[
\sum_{n}b_{n}^{\star}(t)i\hbar\frac{\partial}{\partial t}
b_{n}(t)\nonumber\\
&&+\sum_{n,m}b_{n}^{\star}(t)
\langle n|i\hbar\frac{\partial}{\partial t}|m\rangle
b_{m}(t)-\sum_{n}b_{n}^{\star}(t){\cal E}_{n}(X(t))b_{n}(t)] \}
\end{eqnarray}
where the second term in the action stands for the term
commonly referred to as Berry's phase\cite{berry}(in the 
interpretation of the phase as dynamical~\cite{berry2}) and 
its off-diagonal counter part. The second 
term is defined by
\begin{eqnarray} 
(U(t)^{\dagger}i\hbar\frac{\partial}{\partial t}U(t))_{nm}
&=&\int d^{3}x v^{\star}_{n}(\vec{x},X(t))
i\hbar\frac{\partial}{\partial t}v_{m}(\vec{x},X(t))\nonumber\\
&\equiv& \langle n|i\hbar\frac{\partial}{\partial t}|m\rangle.
\end{eqnarray}

In the operator formulation of the second quantized theory,
we thus obtain the effective Hamiltonian (depending on Bose or 
Fermi statistics)
\begin{eqnarray}
\hat{H}_{eff}(t)&=&\sum_{n}b_{n}^{\dagger}(t)
{\cal E}_{n}(X(t))b_{n}(t)
-\sum_{n,m}b_{n}^{\dagger}(t)
\langle n|i\hbar\frac{\partial}{\partial t}|m\rangle
b_{m}(t)
\end{eqnarray}
with $[b_{n}(t), b^{\dagger}_{m}(t)]_{\mp}=\delta_{n,m}$.
Note that these formulas (4), (8) and (10) are exact. The 
off-diagonal geometric terms in (10), which are 
crucial in the analysis below, are missing in the usual 
adiabatic approximation in the first quantization~\footnote{
It is possible to show that 
\begin{eqnarray}
&&\langle n|T^{\star}\exp\{-(i/\hbar)\int_{0}^{T}dt
\hat{{\cal H}}_{eff}(t)\}
|n\rangle\nonumber\\
&&=\langle n(T)|T^{\star}\exp\{-(i/\hbar)\int_{0}^{T}dt \hat{H}
(\hat{\vec{p}},\hat{\vec{x}},X(t))\}|n(0)\rangle
\nonumber
\end{eqnarray}
where $T^{\star}$ stands for the time ordering operation.
The state $|n\rangle$ on the left-hand side is 
defined by $b_{n}^{\dagger}(0)|0\rangle$ whereas 
$|n(0)\rangle$ and $|n(T)\rangle$ on the right-hand side are 
defined by the eigenfunctions of 
$\hat{H}(\hat{\vec{p}},\hat{\vec{x}},X(t))$.
We defined the Schr\"{o}dinger picture by
\begin{eqnarray}
&&\hat{{\cal H}}_{eff}(t)\equiv 
U(t)^{\dagger}\hat{H}_{eff}(t)U(t)\nonumber\\
&&=\sum_{n}b_{n}^{\dagger}(0)
{\cal E}_{n}(X(t))b_{n}(0)
-\sum_{n,m}b_{n}^{\dagger}(0)
\langle n|i\hbar\frac{\partial}{\partial t}|m\rangle
b_{m}(0)\nonumber
\end{eqnarray}
by introducing $U(t)$, 
$i\hbar\frac{\partial}{\partial t}U(t)= - \hat{H}_{eff}(t)U(t)$,
with $U(0)=1$.}.
In our picture, all the phase factors are included in the 
Hamiltonian, and for this reason, the terminology ``geometric 
terms'' is used for the terms in the Hamiltonian and the ``
geometric phases'' is reserved for the geometric terms when they 
are explicitly interpreted as the phase factors of a specific 
state vector.  

We are mainly interested in the topological properties in the 
infinitesimal neighborhood of level crossing. We thus assume 
that the level crossing takes place only between the lowest two 
levels, and we consider the familiar idealized model with only 
the lowest two levels. This simplification is expected to be 
valid for the analysis of the issues we are interested in.
The effective Hamiltonian to be analyzed 
in the path integral (4) is then defined  by the $2\times 2$
matrix $ h(X(t))=\left(E_{nm}(X(t))\right)$.
If one assumes that the level crossing takes place at the 
origin of the parameter space $X(t)=0$, one needs to analyze
the matrix
\begin{eqnarray}
h(X(t)) = \left(E_{nm}(0)\right) + 
\left(\frac{\partial}{\partial X_{k}}E_{nm}(0)\right) X_{k}(t)
\end{eqnarray}
 for sufficiently small $(X_{1}(1),X_{2}(1), ... )$. By a time 
independent unitary transformation, which does not induce 
a geometric term, the first term is diagonalized.
In the present approximation, essentially the four dimensional 
sub-space of the parameter space is relevant, and after a 
suitable re-definition of the parameters by taking linear 
combinations of  $X_{k}(t)$, we write the matrix as~\cite{berry}
\begin{eqnarray}
h(X(t))
&=&\left(\begin{array}{cc}
            E(0)+y_{0}(t)&0\\
            0&E(0)+y_{0}(t)
            \end{array}\right)
        +g \sigma^{l}y_{l}(t)
\end{eqnarray}
where $\sigma^{l}$ stands for the Pauli matrices, and $g$ is a 
suitable (positive) coupling constant.
 
The above matrix is diagonalized in a standard manner 
\begin{eqnarray} 
h(X(t))v_{\pm}(y)=(E(0)+y_{0}(t) \pm g r)v_{\pm}(y)
\end{eqnarray}
where $r=\sqrt{y^{2}_{1}+y^{2}_{2}+y^{2}_{3}}$  and
\begin{eqnarray}
v_{+}(y)=\left(\begin{array}{c}
            \cos\frac{\theta}{2}e^{-i\varphi}\\
            \sin\frac{\theta}{2}
            \end{array}\right), \ \ \ \ \ 
v_{-}(y)=\left(\begin{array}{c}
            \sin\frac{\theta}{2}e^{-i\varphi}\\
            -\cos\frac{\theta}{2}
            \end{array}\right)
\end{eqnarray}
by using the polar coordinates, 
$y_{1}=r\sin\theta\cos\varphi,\ y_{2}=r\sin\theta\sin\varphi,
\ y_{3}=r\cos\theta$. Note that
$v_{\pm}(y(0))=v_{\pm}(y(T))$ if $y(0)=y(T)$ except for 
$(y_{1}, y_{2}, y_{3}) = (0,0,0)$, and $\theta=0\ {\rm or}\ \pi$.
If one defines
\begin{eqnarray} 
v^{\dagger}_{m}(y)i\frac{\partial}{\partial t}v_{n}(y)
=A_{mn}^{k}(y)\dot{y}_{k}
\end{eqnarray}
where $m$ and $n$ run over $\pm$,
we have
\begin{eqnarray}
A_{++}^{k}(y)\dot{y}_{k}
&=&\frac{(1+\cos\theta)}{2}\dot{\varphi},
\nonumber\\
A_{+-}^{k}(y)\dot{y}_{k}
&=&\frac{\sin\theta}{2}\dot{\varphi}+\frac{i}{2}\dot{\theta}
=(A_{-+}^{k}(y)\dot{y}_{k})^{\star}
,\nonumber\\
A_{--}^{k}(y)\dot{y}_{k}
&=&\frac{1-\cos\theta}{2}\dot{\varphi}.
\end{eqnarray}
The effective Hamiltonian (10) is then given by 
\begin{eqnarray}
&&\hat{H}_{eff}(t)=(E(0)+y_{0}(t) + g r(t))b^{\dagger}_{+}b_{+}
\nonumber\\
&&+(E(0)+y_{0}(t) - g r(t))b^{\dagger}_{-}b_{-}
 -\hbar \sum_{m,n}b^{\dagger}_{m}A^{k}_{mn}(y)\dot{y}_{k}b_{n}.
\end{eqnarray}

In the conventional adiabatic approximation, one approximates
the effective Hamiltonian (17) by
\begin{eqnarray}
\hat{H}_{eff}(t)&\simeq& (E(0)+y_{0}(t) + g r(t))
b^{\dagger}_{+}b_{+}\nonumber\\
&&+(E(0)+y_{0}(t) - g r(t))b^{\dagger}_{-}b_{-}\nonumber\\
&&-\hbar [b^{\dagger}_{+}A^{k}_{++}(y)\dot{y}_{k}b_{+}
+b^{\dagger}_{-}A^{k}_{--}(y)\dot{y}_{k}b_{-}]
\end{eqnarray}
which is valid for $Tg r(t)\gg \hbar\pi$, the magnitude of the 
geometric term.
The Hamiltonian for $b_{-}$, for example, is then eliminated by 
a ``gauge transformation''
\begin{eqnarray}
&&b_{-}(t)=\exp\{-(i/\hbar)\int_{0}^{t}dt[
E(0)+y_{0}(t) - g r(t) 
-\hbar A^{k}_{--}(y)\dot{y}_{k}] \} \tilde{b}_{-}(t)
\end{eqnarray}
in the path integral (8), and the amplitude 
$\langle 0|\hat{\psi}(T)b^{\dagger}_{-}(0)|0\rangle$, which 
corresponds to the probability amplitude in the first 
quantization, is given by (up to a wave function 
$\phi_{E}(\vec{x})$)
\begin{eqnarray}
&&\exp\{-\frac{i}{\hbar}\int_{0}^{T}dt[
E(0)+y_{0}(t) - g r(t) 
-\hbar A^{k}_{--}(y)\dot{y}_{k}] \}
\nonumber\\
&&\times v_{-}(y(T))
\langle 0|\tilde{b}_{-}(T)\tilde{b}^{\dagger}_{-}(0)|0\rangle
\end{eqnarray}
with $\langle 0|\tilde{b}_{-}(T)\tilde{b}^{\dagger}_{-}(0)
|0\rangle=1$.
For a $2\pi$ rotation in $\varphi$ with fixed $\theta$, for 
example, the geometric term  gives rise to the well-known factor
$\exp\{i\pi(1-\cos\theta) \}$ by using (16)~\cite{berry}. The 
corrections to the phase due to the finite $1/T$ may be 
evaluated by an iterative procedure~\cite{berry2}, for example.

Another representation, which is useful to analyze the behavior
near the level crossing point, is obtained by a further unitary 
transformation
$b_{m}=U(\theta(t))_{mn}c_{n}$ where $m,n$ run over $\pm$
with
\begin{equation}
U(\theta(t))=\left(\begin{array}{cc}
            \cos\frac{\theta}{2}&-\sin\frac{\theta}{2}\\
            \sin\frac{\theta}{2}&\cos\frac{\theta}{2}
            \end{array}\right),
\end{equation}
and the above effective Hamiltonian (17) is written as
\begin{eqnarray}
\hat{H}_{eff}(t)&&= (E(0)+y_{0}(t)+gr\cos\theta)
c^{\dagger}_{+}c_{+}\nonumber\\
&&+(E(0)+y_{0}(t)-gr\cos\theta)c^{\dagger}_{-}c_{-}\nonumber\\
&&-gr\sin\theta c^{\dagger}_{+}c_{-}
-gr\sin\theta c^{\dagger}_{-}c_{+}
-\hbar\dot{\varphi} c^{\dagger}_{+}c_{+}.
\end{eqnarray}
In the above unitary transformation, an extra geometric
term $-U(\theta)^{\dagger}i\hbar\partial_{t}U(\theta)$ is 
induced by the kinetic term of the path integral 
representation (8). One can
confirm that this extra term precisely cancels the term 
containing $\dot{\theta}$ in $b^{\dagger}_{m}
A^{k}_{mn}(y)\dot{y}_{k}b_{n}$ as in (16). 
We thus diagonalize the geometric terms in this representation.
We also note that 
$U(\theta(T))=U(\theta(0))$ if $X(T)=X(0)$ except for the 
origin, and thus the initial and final states receive the same 
transformation in scattering amplitudes. 

In the infinitesimal neighborhood of the level crossing point,
namely, for sufficiently close to the origin of the parameter 
space $(y_{1}(t), y_{2}(t), y_{3}(t) )$ but 
$(y_{1}(t), y_{2}(t), y_{3}(t))\neq (0,0,0)$, one may 
approximate (22) by
\begin{eqnarray}
&&\hat{H}_{eff}(t)\simeq (E(0)+y_{0}(t)+gr\cos\theta)
c^{\dagger}_{+}c_{+}\nonumber\\
&&+(E(0)+y_{0}(t)-gr\cos\theta)c^{\dagger}_{-}c_{-}
-\hbar\dot{\varphi} c^{\dagger}_{+}c_{+}.
\end{eqnarray}
To be precise, for any given {\em fixed} time interval $T$,
$T\hbar\dot{\varphi}\sim 2\pi\hbar$
which is invariant under the uniform scale transformation 
$y_{k}(t)\rightarrow 
\epsilon y_{k}(t)$. On the other hand, 
one has $T gr\sin\theta \rightarrow T\epsilon gr\sin\theta$
by the above scaling, and thus one can choose 
$T\epsilon gr\ll \hbar$. The terms $\pm gr\cos\theta$ in (23)
may also be ignored in the present approximation.

In this new basis (23), the geometric phase appears only for
 the mode $c_{+}$ which gives rise to a phase factor
$\exp\{i\int_{C} \dot{\varphi}dt \}=\exp\{2i\pi \}=1$,
and thus no physical effects. In the infinitesimal neighborhood 
of level crossing, the states spanned by 
$(b_{+},b_{-})$ are transformed to a linear combination of the 
states spanned by $(c_{+},c_{-})$, which give no non-trivial 
geometric phases. The geometric terms are topological 
in the sense that they are invariant under the uniform scaling 
of 
$y_{k}(t)$, but their physical implications in conjunction with 
other terms in the effective Hamiltonian are not. For example, 
starting with the state
$b^{\dagger}_{-}(0)|0\rangle$ one may first make 
$r\rightarrow  small$ with fixed $\theta$ and $\varphi$, 
then make a $2\pi$ rotation in $\varphi$ in the bases 
$c^{\dagger}_{\pm}|0\rangle$, and then come back to 
the original $r$ with fixed $\theta$ and $\varphi$ for a given
fixed  $T$; in this cycle, one does not pick up any non-trivial 
geometric phase even though one covers the solid angle 
$2\pi(1-\cos\theta)$. The transformation from $b_{\pm}$ to 
$c_{\pm}$ is highly non-perturbative, since a complete 
rearrangement of two levels is involved.

It is noted that one cannot simultaneously diagonalize the 
conventional energy eigenvalues and the induced 
geometric terms in (17) which is exact in the present two-level
model (12). The topological 
considerations~\cite{stone, berry} are thus inevitably 
approximate. In this respect, it may be instructive to 
consider a model without level crossing which is defined by 
setting 
\begin{eqnarray}
y_{3}=\Delta E/2g 
\end{eqnarray}
in (17), 
where $\Delta E$ stands for the minimum of the level spacing. 
The geometric terms then loose invariance under the uniform 
scaling of $y_{1}$ and $y_{2}$.
In the limit 
\begin{eqnarray}
\sqrt{y^{2}_{1}+y^{2}_{2}}\gg\Delta E/2g,
\end{eqnarray}
(and thus $\theta\rightarrow \pi/2$), the geometric terms in (17) 
exhibit approximately topological behavior for the reduced 
variables $(y_{1},y_{2})$. Near the point where the level 
spacing becomes minimum, which is specified by 
\begin{eqnarray}
(y_{1},y_{2}) \rightarrow (0,0)
\end{eqnarray}
(and thus $\theta\rightarrow0$), the 
geometric terms in (17) assume the same form of the geometric 
term as in (23). Our analysis above shows that the model with 
level crossing exhibits precisely the same topological 
properties for  any finite $T$. 

It is instructive to analyze an explicit 
example in Refs.~\cite{geller,bhandari} where the following 
parametrization has been introduced
\begin{eqnarray}
(y_{1},y_{2},y_{3})=(B_{0}(b_{1}+\cos\omega t), 
B_{0}\sin\omega t, B_{z})
\end{eqnarray}
and $g=\mu$.  The case $b_{1}=0$ and $B_{z}\neq 0$ corresponds to
the model without level crossing discussed above, and the 
geometric phase becomes trivial for $B_{0}\rightarrow 0$. The 
case $b_{1}=B_{z} = 0$ describes the situation in (23), namely,
a closed cycle in the infinitesimal neighborhood of level 
crossing for $B_{0}\rightarrow 0$ with $T=2\pi/\omega$ kept 
fixed,
and the geometric phase becomes trivial. On the other hand,
the usual adiabatic approximation (18) (with $\theta=\pi/2$ in
the present model) in the neighborhood of level crossing is 
described by $b_{1}=B_{z} = 0$ and $B_{0}\rightarrow 0$ (and
$\omega\rightarrow 0$) with 
\begin{eqnarray}
\mu B_{0}/\hbar\omega\gg 1 
\end{eqnarray}
kept fixed, namely, the effective magnetic field is always 
strong; the topological proof of 
phase-change rule~\cite{stone} is based on the consideration of 
this case. It should be noted that the geometric phase 
becomes trivial for $B_{0}\rightarrow 0$ with $b_{1}=B_{z} = 0$ 
and 
\begin{eqnarray}
\mu B_{0}/\hbar\omega\ll 1 
\end{eqnarray}
kept fixed. (If one starts with $b_{1}=B_{z} = 0$ and 
$\omega=0$, of course, no geometric terms.) 
It is clear that the topology is non-trivial only for a quite 
narrow (essentially measure zero) window of the parameter space
 $(B_{0}, \omega)$ in the approach to the level crossing
$B_{0}\rightarrow 0$. In 
this analysis, it is important to distinguish the level 
crossing problem from the motion of a spin $1/2$ particle; the 
wave functions (14) are single valued for a $2\pi$ rotation in 
$\varphi$ with fixed $\theta$.  

The path integral (4), where the 
Hamiltonian is diagonalized both at $t=0$ and $t=T$ if 
$X(T)=X(0)$, shows no obvious singular behavior at the level 
crossing point.
On the other hand, the path integral (8) is  subtle at the 
level crossing point; the bases $\{v_{n}(\vec{x},X(t))\}$ are 
singular on top of level crossing as in (14), and thus the 
unitary transformation $U$ to (8) and the induced geometric 
terms become singular there. The present analysis suggests that 
 the path integral is not singular for any finite $T$, as is 
expected from (4). We consider that this result is natural since
 the starting Hamiltonian (1) does not contain any obvious 
singularity.

The conventional treatment of geometric phases in the (precise) adiabatic approximation is based on
 the premise that one 
can choose $T$ sufficiently large for any {\em given} 
$\epsilon\sim r$ such that 
\begin{eqnarray}
Tg\epsilon \gg \hbar,
\end{eqnarray}
and thus $T\rightarrow \infty$ for $\epsilon\rightarrow 0$, 
namely, it takes an infinite amount of time to approach the 
level crossing point~\cite{berry, simon}.
Finite $T$ may however be appropriate in
practical applications, as is noted in~\cite{berry}. 
Because of the uncertainty principle
$T\Delta E \geq \frac{1}{2}\hbar$,
the (physically measured) energy uncertainty for any given fixed 
$T$ is not much different from the magnitude of the geometric 
term $2\pi\hbar$, and the level spacing becomes much smaller 
than these values in the infinitesimal neighborhood of level 
crossing for the given $T$. An intuitive picture behind (23) is
that the motion in  $\dot{\varphi}$ smears the 
``monopole'' singularity for 
arbitrarily large but finite $T$. 

The notion of Berry's phase is known to be useful in various 
physical contexts~\cite{shapere}-\cite{review}, and the 
topological considerations are often crucial to obtain a 
qualitative understanding of what is going on. Our analysis 
however shows that the geometric phase associated with level 
crossing becomes topologically trivial  in practical physical 
settings with any finite $T$. This is in sharp contrast to the 
Aharonov-Bohm phase~\cite{aharonov} which is induced by the 
time-independent gauge potential and topologically exact for 
any finite time interval $T$. The similarity and difference 
between the geometric phase associated with level crossing 
and the Aharonov-Bohm phase have been recognized in the early
literature~\cite{berry, aharonov}. Also, the correction to the
geometric phase in terms of the small slowness parameter 
$\epsilon$ has been analyzed, and the closer to a 
degeneracy a system passes the slower is the necessary passage 
for adiabaticity has been noted in~\cite{berry2}. But,
to our knowledge, the fact that the geometric phase becomes
topologically trivial for practical physical settings 
with any fixed finite $T$, such as in the practical 
Born-Oppenheimer 
approximation where a large but finite ratio of two time 
scales is involved, has not been clearly stated in the 
literature. We emphasize that this fact is proved independently 
of the adiabatic approximation.

The notion of the geometric phase is useful, but great care 
needs to be exercised as to its topological properties.

\end{document}